\documentclass[useAMS,usenatbib,fleqn,twocolumn]{mnras}
\usepackage[normalem]{ulem}
\usepackage[utf8]{inputenc}
\usepackage{enumerate}
\usepackage{amsmath, amssymb,  graphicx, epsfig, epsfig, slashed}
\usepackage[mathscr]{euscript}
\let\oldcdot\cdot
\usepackage{breqn}
\let\cdot\oldcdot
\usepackage{epsfig}  
\usepackage{graphicx}   
\usepackage{slashed}       
\usepackage{tikz}
\usepackage{subcaption} 
\usepackage{booktabs}
\usepackage{url}
\usepackage{color,soul} 
\usepackage{xcolor}
\usepackage{multirow} 
\usepackage{mathtools}
\usepackage{bm}
\usepackage{hyperref}
\usepackage[all]{hypcap}
\usepackage{ulem}
\hypersetup{  
    colorlinks=true,
    linkcolor=blue,
    filecolor=red,      
    urlcolor=cyan,
    citecolor=blue,
    }

\usepackage[T1]{fontenc}
\usepackage{ae,aecompl}
\let\oldcdot\cdot
\usepackage{breqn}
\let\oldcdot\cdot

\newcommand{\be}{\begin{eqnarray}}
\newcommand{\ee}{\end{eqnarray}}
\newcommand{\f}{\Tilde{f}}

\newcommand{\deltat}{\overline{T}_{21}}

\newcommand{\tk}{T_{\rm K}}
\newcommand{\tgamma}{T_{\gamma}}
\newcommand{\cint}{C_{\rm int}}

\def \TBRIGHT{T_{\rm{21}}}

\def \YA{\bar{Y}_{\alpha}}
\def\spose#1{\hbox to 0pt{#1\hss}}

\title[Co-SIMP model with dark ages 21-cm signal]{Exploring the Co-SIMP dark matter model using the 21-cm signal from the dark ages}

\author[Paul et al.]{Debarun Paul,$^{1}$\thanks{debarun31paul@gmail.com} 
Sourav Pal,$^{1}$ Deepthi Moorkanat,$^{2}$ Antara Dey,$^{1}$ Amit Dutta Banik,$^{1}$
\newauthor Rajesh Mondal$^{2}$\\
\\
$^{1}$ Physics and Applied Mathematics Unit, Indian Statistical Institute, 203 B.T. Road, Kolkata 700108, India\\
$^{2}$ Department of Physics, National Institute of Technology Calicut, Calicut 673601, Kerala, India}

\date{}

\begin{document}
\label{firstpage}
\pagerange{\pageref{firstpage}--\pageref{lastpage}}
\maketitle
\thispagestyle{empty} 

\begin{abstract}
The redshifted 21-cm signal from the dark ages offers a powerful probe of cosmological models and the underlying dark matter (DM) microphysics. We investigate deviations from the standard $\Lambda$CDM prediction, an absorption trough of approximately $-40.6\,\mathrm{mK}$ at redshift $z\simeq85.6$, in the context of co-SIMP (strongly interacting massive particle) DM. The co-SIMP interaction strength is encoded by the parameter $C_{\rm int}$, incorporating the masses of DM and standard model (SM) particles, the interaction cross-section, and the amount of heat exchange between the two sectors. Increasing $C_{\rm int}$ deepens the absorption feature and shifts the trough to higher redshifts in the global signal.
For $C_{\rm int}=1.0$, the minimum brightness temperature reaches $-50.6,\mathrm{mK}$ at $z\simeq86.2$. The 21-cm power spectrum increases with $C_{\rm int}$ in addition to the global signal.
We assess the detectability of these signatures using signal-to-noise ratio (SNR) and Fisher forecasts. The maximum SNR reaches $\sim 15.7$ for $C_{\rm int}=1.0$ for the global signal. Fisher forecasts for $1,000$ hours of integration time show that this model can be distinguished from a null-signal at $4.3\sigma$ and a mild 1.6$\sigma$ from $\Lambda$CDM, improving by an order of magnitude for 100,000 hours. For the 21-cm power spectrum, a $5,\mathrm{km}^2$ array with 1,000 hours yields a $4.63\sigma$ detection and mildly separated from the standard scenario at $1.78\sigma$. These findings highlight the potential of the 21-cm cosmology to probe the properties of DM and demonstrate that upcoming dark ages experiments, particularly space-based and lunar observations, can offer a promising avenue to test co-SIMP models.

\end{abstract}




\section{Introduction}
\label{sec:intro}
The dark ages, spanning roughly from the epoch of recombination (redshift $z \sim 1100$) down to the formation of the first luminous objects ($z \sim 30$), are a special epoch in the history of the Universe. During this period, the Universe was not yet complicated by astrophysical processes. This makes it a good probe of fundamental cosmology, novel dark matter theories, and perhaps fundamental physics. We can study the dark ages using the redshifted 21-cm signal from neutral hydrogen (HI) \citep{Sunyaev1972, Scott1990}. It allows us to create a map of the early Universe over a wide period. This is because the different frequencies we observe tell us about different times. So, the 21-cm signal gives us a three-dimensional (3D) view. This is different from the cosmic microwave background (CMB), which is two-dimensional (2D). We can study the global 21-cm signal (mean) or how it changes from place to place (its power spectrum). By doing this, we can learn much more about cosmology than the CMB alone. This opens up a new way to look at the early Universe \citep{loeb2004, Mondal:2023xjx,Saha:2021pqf}.

The temperature of the CMB ($T_\gamma$) decreased as $(1+z)$. During the dark ages, the kinetic temperature of the gas ($T_{\rm K}$) was adiabatically cooled as $(1+z)^2$, resulting $T_{\rm K}<T_{\gamma}$. Another important temperature is the spin temperature ($T_{\rm S}$), which tells us about the ratio of the population densities of the upper and lower states of the 21-cm transition of HI. Collisions between atoms kept $T_{\rm S}$ strongly coupled to $T_{\rm K}$ until a redshift of about $z \sim 70$, resulting $T_{\rm S}<T_{\gamma}$. After this time, since the gas density started to decrease due to the expansion of the Universe, these collisions started to became less effective, which results that $T_{\rm S}$ started to approach $T_\gamma$. At redshift of about $z\sim 30$, the collision couplings became completely ineffective and as a result, $T_{\rm S}$ became completely coupled to $T_{\gamma}$. Therefore, for an extended period in the dark ages, specifically between redshifts of about $z \sim 200$ and $z \sim 30$, $T_{\rm S}$ was much lower than $T_\gamma$. Because $T_{\rm S}$ was lower than $T_\gamma$, the neutral hydrogen atoms absorbed the photons from the CMB. This absorption is what we expect to see as the 21-cm signal from the dark ages.

The 21-cm power spectrum in the dark ages is sensitive to cosmological parameters, especially on the scale of baryon acoustic oscillations (BAOs) \citep{barkana2005, Mondal:2023xjx,Park:2025phj}. Since, during the dark ages, there was no complicated astrophysics and fluctuations were mostly linear, this makes modelling and understanding them easier than in the later Universe. Fluctuations in baryon density, peculiar velocity, and baryon temperature are imprinted in the 21-cm power spectrum \citep{Bharadwaj_2004, Barkana:2004zy, Naoz2005, barkana2005}. Smaller effects, \textit{e.g.} impacts of supersonic relative velocities of standard model (SM) and dark matter (DM) at all the scales of the matter power spectra, full linear analysis of the perturbed 21-cm optical depth and perturbed recombination effects on gas temperature, must also be considered for accurate calculations, as stated in \cite{Ali-Ha2014, Lewis2007}. It has also been shown that the 21-cm signal could be a strong way to study primordial non-Gaussianity~\citep{Munoz:2015eqa, Pillepich:2006fj,Yamauchi:2022fri}. However, observing the high wavenumbers where this is the most promising would be challenging.

Observing the 21-cm signal from the dark ages requires very low frequencies, below 45 MHz. The ionosphere severely distorts and blocks these radio waves, making Earth-based observations nearly impossible. Therefore, lunar or space-based radio telescopes are necessary. Several international missions are being developed for this purpose as part of the race to return to the Moon. These include NCLE\footnote{\url{https://doi.org/10.1126/science.aau2004}} (Netherlands–China), DAPPER (USA; \citealt{dapper}), FARSIDE (USA; \citealt{farside}), PRATUSH\footnote{\url{https://wwws.rri.res.in/DISTORTION/pratush.html}} (India), FarView (USA; \citealt{dapper}), SEAMS (India, \citealt{seams}), LuSee Night (USA; \citealt{luseenight}), ALO\footnote{\url{https://www.astron.nl/dailyimage/main.php?date=20220131}} (Europe), ROLSES\footnote{\url{https://www.colorado.edu/ness/ness-projects}} (USA), and Tsukuyomi\footnote{\url{https://www.sankei.com/article/20240517-5ZRIGLFVXJP4RA4CRVHJBTDED4}} (Japan). In addition, space-based missions have also been proposed, including CoDEX~\citep{2021ExA....51.1641K} and DSL/Hongmeng project\footnote{\url{https://www.astron.nl/dsl2015}} (China; \citealt{2019arXiv190710853C}). Collectively, these missions aim to study either the global signal or the spatial variations of the 21-cm signal from the dark ages. Although most of these experiments are in the early design stages, measuring the dark ages power spectrum is a more distant goal than measuring the global signal. The Moon offers practical benefits beyond avoiding the ionosphere, such as a dry, stable environment and shielding from radio interference from Earth, especially on the far side.

The 21-cm signal from the dark ages holds great promise for precise cosmology, and new observations are developing quickly. A study by \cite{Mondal:2023xjx} explored how this signal can be used to measure cosmological parameters. Their work showed that observing the global 21-cm signal for 1,000\,hours could allow us to measure certain combinations of cosmological parameters with an accuracy of about 10\%. Measuring 21-cm fluctuations is more challenging. It would require a very large collecting area, around 10\,km$^2$. This is bigger than any current radio telescope, but it might be possible for future instruments. With such a large area and 1,000\,hours of observation, the measurement accuracy could be twice as good as with the global signal. Therefore, if we assume our current understanding of cosmology is correct, 21-cm observations from the dark ages could significantly advance our knowledge of the Universe.

The discussion so far has assumed standard cosmology. However, thinking about new discoveries that might lie beyond our current understanding is also important. There could be non-standard models that other cosmic observations allow. These models might be detectable with 21-cm observations of the dark ages. It is generally not easy to propose such models, given the existing limits from other cosmic observations (\textit{e.g.}, CMB) and studies of dark matter. Interestingly, a possible detection by the EDGES (Experiment to Detect the Global Epoch of Reionization Signature) experiment \citep{Bowman:2018yin} of a strong 21-cm signal during the cosmic dawn brings about several such theories. Although the SARAS (Shaped Antenna measurement of the background RAdio Spectrum) experiment has questioned this EDGES result \citep{SARAS}, and more measurements are needed to clarify the situation, the initial EDGES claim has inspired theories with a wide range of possibilities. These new theories can be tested, regardless of whether the EDGES signal turns out to be correct. \cite{Mondal:2023bxb} recently studied two such models. The first model was a uniform extra radio background (ERB) with a synchrotron spectrum. This spectrum is interesting because the ERB might explain some or all of the observed radio background from outside our galaxy, as noted by \citep{fialkov19}. The second model they considered was millicharged dark matter (mDM), proposed by \citep{munoz18}. In this model, a small portion of dark matter particles have a very small electric charge.

The standard DM models presumably consider a $\mathbb{Z}_2$ symmetry to stabilise the WIMP (weakly interacting massive particle) dark matter~\citep{silveira1985scalar,mcdonald1994gauge,burgess2001minimal}. While minimal and phenomenologically successful, this framework suffers certain limitations. $\mathbb{Z}_2$ symmetry restricts DM number-changing processes to even-number interactions, leaving a restricted freedom in achieving the observed relic abundance of DM. Also, in the scenario where the DM annihilation process is suppressed, $\mathbb{Z}_2$ symmetry provides no additional natural mechanisms to prevent DM-overabundance. Additionally, the indirect bounds on the pair annihilations are often severe~\citep{HESS:2021zzm,HESS:2015cda,HESS:2014zqa,Thorpe-Morgan:2020czg,Fermi-LAT:2016uux,Fermi-LAT:2015att}. On the other hand, models with higher discrete symmetries, such as $\mathbb{Z}_3$ offer richer dynamics, providing additional freedom to achieve the observed relic abundance of DM. Furthermore, $\mathbb{Z}_3$ models have the ability to address small scale structure issues of the Universe~\citep{burkert1995structure,moore1999dark,boylan2011too,oman2015unexpected,Parikh:2023qtk}, offering a potential resolution of the key limitations of the $\Lambda$CDM ($\Lambda$ cold dark matter) scenario. In this paper, we consider the $\mathbb{Z}_3$ symmetric co-SIMP (strongly interacting massive particle) model, having a distinctive freeze-out mechanism from a conventional WIMP \citep{Smirnov:2020zwf,Parikh:2023qtk}. While the annihilating DM heats up the intergalactic medium (IGM), the $\mathbb{Z}_3$ symmetric co-SIMP dark matter with its unique interaction property leading to endothermic reactions with IGM leaves its imprint by influencing the 21-cm signal from the dark ages. This enables co-SIMP DM to provide a feasible explanation for the dip of the EDGES signal, as demonstrated in \cite{Paul:2023emc}, in addition to addressing the small-scale structure issues of Universe. 
These characteristics of the co-SIMP DM model make it a promising DM candidate from both a cosmological and a phenomenological perspective. The prospect of probing the co-SIMP process through various experiments, such as beam dump searches \citep{Marsicano:2018vin,Batell:2014mga} and electron $g-2$ measurements \citep{Muong-2:2021ojo}, further enriches interest in investigating different aspects of this model. We explore the cosmological implications of the co-SIMP scenario during the dark ages, focusing on its imprint on the global 21-cm signal, power spectrum, and the associated observational prospects.

This paper is organized as follows. Section~\ref{sec:DMmodel} explains the framework of the co-SIMP dark matter model. The effect of this model on the global 21-cm signal is discussed in Section~\ref{sec:darkage_global}. Section~\ref{sec:darkage_ps} describes how this model affects the 21-cm power spectrum during the dark ages. The possibility of finding evidence for this model and distinguishing it from the standard $\Lambda$CDM model is discussed in Section~\ref{sec:prospects}. Lastly, Section~\ref{sec:conclusion} provides a summary and discussion of our results. 

Throughout our analysis, we adopt the Hubble parameter, $H_0=67.81~{\rm km\,s}^{-1}{\rm Mpc}^{-1}$, the present-day total matter density, $\Omega_{\rm m}^{(0)} h^2=0.14175$ and the present-day baryonic density $\Omega_{\rm b}^{(0)} h^2=0.02245$, as inferred from the statistical analysis of the co-SIMP model using Planck-18+BAO dataset \citep{Paul:2023emc}~\footnote{In our earlier analysis \citep{Paul:2023emc}, we performed a detailed Markov chain Monte Carlo (MCMC) analysis to constrain the cosmological parameters within the co-SIMP dark matter framework using the combined Planck-18+BAO dataset. Since the present work is consistently formulated within the co-SIMP scenario, employing cosmological parameters inferred for this model is an internally consistent choice, rather than adopting those of the standard $\Lambda$CDM model. The best-fit parameters were found to lie within $1\sigma$ of the Planck-18+BAO $\Lambda$CDM values, resulting in only minor changes in the matter power spectra, which do not qualitatively affect the final conclusion.}.

\section{Framework of co-SIMP dark matter model}
\label{sec:DMmodel}
A compelling alternative to the WIMP paradigm is the class of SIMPs, where the abundance of DM relics is governed by number-changing processes, rather than simple annihilation into SM particles. In this work, we focus on a particular subclass known as co-SIMP \citep{Smirnov:2020zwf,Parikh:2023qtk}. In this scenario, the DM particles ($\chi$) are stabilized by a discrete $\mathbb{Z}_3$ symmetry and participate in $2\to3$ interactions involving a SM particle ($\psi$). The defining interaction of the co-SIMP framework is given by
\begin{eqnarray}
\label{eq:cosimp_int}
    \chi+\psi \rightarrow \chi+\chi+\psi,
\end{eqnarray}
with $M_{\chi}$ and $M_{\psi}$ being the mass of the DM and SM particles. This process alters the DM number density, setting the relic abundance via interactions that do not deplete DM into purely SM states. The realization of minimal UV complete particle physics of the co-SIMP framework features a Dirac fermion DM particle $\chi$, which is charged under a dark $U(1)_D$ gauge symmetry \citep{Parikh:2023qtk}. A residual $\mathbb{Z}_3$ symmetry arising from the spontaneous breaking of $U(1)_D$ stabilizes the dark matter candidate and mixing of dark photon with ordinary photon contributes to the effective number of relativistic species ($N_{\rm eff}$), further restraining co-SIMP DM parameters (interaction strength) with Planck-18 experiments (see \cite{Paul:2023emc} for details). The kinetic mixing of dark photons facilitates the crucial co-SIMP process and allows the model to be probed by $g-2$ and beam dump experiments.

\begin{figure}[]
    \centering
    \includegraphics[scale=0.3]{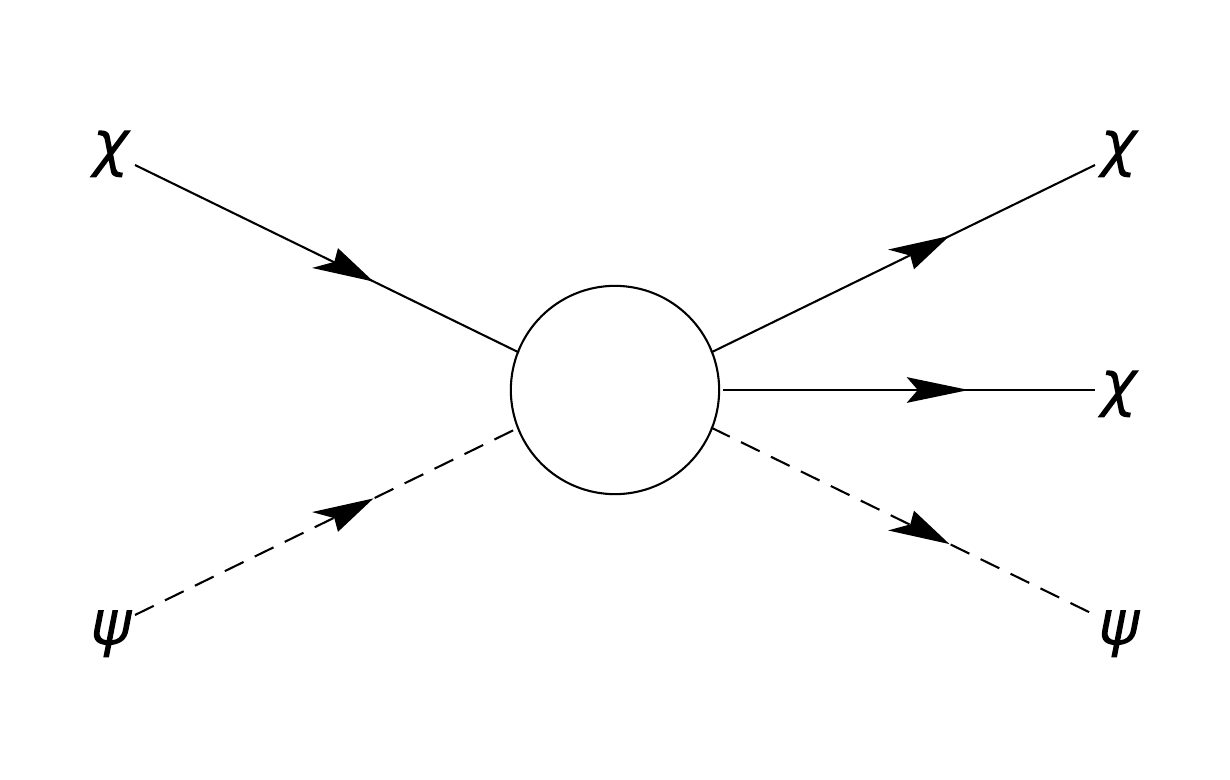}
    \caption{Feynman diagram diagram for $2\rightarrow 3$ co-SIMP interaction process. $\chi$ and $\psi$ represent DM and SM particles, respectively.}
    \label{fig:cosimp}
\end{figure}

Fig.~\ref{fig:cosimp} schematically represents the Feynman diagram for this $2\to3$ interaction. In this work, we specifically consider the leptonic interaction, in which $\psi$ is identified with the electron, with possible allowed interaction terms $y_{\phi e}\phi\bar{e}e$ and $y_{\phi\chi}\phi\chi^c\chi$ \citep{Parikh:2023qtk}.

A crucial feature of this setup is the kinematic constraint $M_{\chi}<2M_{\psi}$, which renders the $2\to2$ annihilation process $\chi+\psi\rightarrow\psi+\psi$ forbidden. As a result, the co-SIMP channel becomes the dominant mechanism shaping the thermal history of $\chi$, avoiding the standard WIMP-like freeze-out. This has two significant consequences. First, DM decouples from the thermal bath via a non-traditional freeze-out, in which freeze-out occurs through $2\to3$ interactions. Second, the leptophilic interaction, particularly for electrons, influences the evolution of the IGM temperature. Due to this direct coupling with the electrons, the model can significantly impact the 21-cm signal. The modified thermal history of the IGM can leave imprints on the differential brightness temperature of the 21-cm line of neutral hydrogen, which is sensitive to the baryon temperature and ionization state. \cite{Paul:2023emc} showed that this interaction can lead to enhanced cooling of the baryonic gas and can naturally explain the absorption dip of the 21-cm signal, as reported by the EDGES collaboration. In this work, we will explore the comprehensive impact of the co-SIMP model on the 21-cm signal during the dark ages.

\section{Signature on the 21-cm global signal}
\label{sec:darkage_global}
In 21-cm experiments, the differential brightness temperature ($T_{21}$) is the primary physical quantity observed. If we consider that the optical depth of the 21-cm transition is significantly less than one, then the sky-averaged (global) brightness temperature during the dark ages can be expressed as \citep{FURLANETTO2006181,2012RPPh...75h6901P,Mondal:2023xjx}: 
\begin{multline}
\label{eq:dTb}
\deltat \simeq 54.0\,{\rm mK}\, \frac{\rho_{\rm HI}}{\bar{\rho}_{\rm H}} \left(\frac{\Omega^{(0)}_{\rm b}h^2}{0.02242}\right) \left(\frac{\Omega^{(0)}_{\rm m}h^2}{0.1424}\right)^{-\frac{1}{2}} \left(\frac{1 + z}{40}\right)^{\frac{1}{2}} \\
\frac{x_{\rm c}}{1 + x_{\rm c}} \left(1 - \frac{T_{\gamma}}{\tk}\right)\ ,
\end{multline}
where $\rho_{\rm HI}$ is the neutral hydrogen density, $\bar{\rho}_{\rm H}$ is the mean hydrogen density, and $x_{\rm c}$ is the collisional coupling coefficient \citep{Kuhlen_2006,PhysRevD.74.103502}. The evolution of $\tk$ with respect to $z$, can be presented as \citep{PhysRevD.74.103502}
\begin{eqnarray}
\label{eq:Tk}
\frac{d\tk}{dz} = \frac{2\tk}{1+z} - \frac{2}{3H(z)(1+z)}\sum_i \frac{\epsilon_i}{k_{\!_{B}} n_{\rm tot}},
\end{eqnarray}
with $n_{\rm tot}$ being the total baryonic number density and the others have their standard meaning. Here, the first part of the expression represents the adiabatic cooling resulting from cosmic expansion, while each subsequent $\epsilon_i$ term in the second part, corresponds to the energy injection or extraction rate per unit volume associated with the $i^{\rm th}$ process. During the dark ages, $\tk$ is mainly affected due to the Compton scattering between electron and photon in the vanilla $\Lambda$CDM scenario, which can be manifested as \citep{Bharadwaj_2004} 
\be \label{eq:epsComp}
\epsilon_{\rm comp} = \frac{3}{2}n_{\rm tot} k_{\!_{B}} \frac{x_{\rm e}}{1 + x_{\rm e} + f_{\!_{\rm He}}}\frac{8 \sigma_{\!_{T}} u_{\gamma}}{3m_e c}\left(\tgamma - \tk\right)\:.
\ee
Here, $\sigma_{\!_{T}}$ and $u_{\gamma}$ respectively represent the Thomson cross-section and the background photon energy density, and the others have their standard meaning. $f_{\!_{\rm He}}\approx 0.08$ is the helium fraction that can be calculated from the helium abundance in the Universe. $x_{\rm e}$ is the total electron fraction, whose redshift evolution is discussed in Appendix~\ref{app:xe}.

\begin{figure}
    \centering
    \includegraphics[scale=0.6]{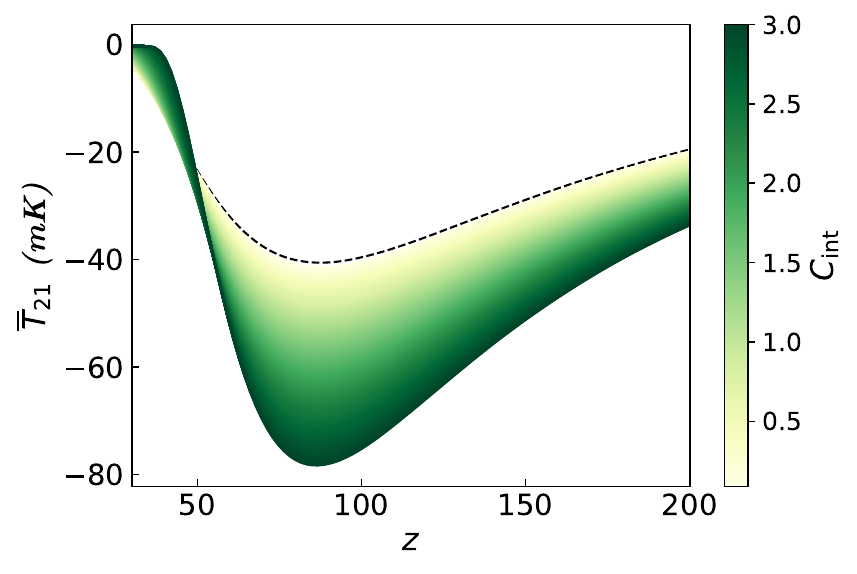}
    \caption{The global signal shows an overall depletion relative to the $\Lambda$CDM  (shown by black dashed line) across entire redshift range. However, for $z \geq 50$, the magnitude of depletion increases with increasing $\cint$, whereas for $z < 50$, the trend reverses, leading to a reduction in depletion with higher $\cint$.}
    \label{fig:global}
\end{figure}

The phenomenology of co-SIMP DM allows the possibility of cooling the IGM temperature by exchanging heat from the standard sector to the dark sector. The rate of that can be expressed, in terms of species density ($\rho$), as \citep{Paul:2023emc}
\begin{eqnarray}\label{eq:2to3_dedvdt}
\left|\epsilon_{\rm co-SIMP}\right|\,&\equiv&\,\left.\frac{dE}{dVdt}\right|_{\rm co-SIMP} \nonumber\\
&=& -\f\, \sqrt{\frac{M_{\chi}c^2}{\left(M_{\psi}c^2\right)^3}}\,\sqrt{\rho_{\psi}^3 \rho_{\chi}}\,\langle \sigma v\rangle_{\rm co-SIMP}.
\end{eqnarray}
Here, $\f$ quantifies the amount of heat exchange between the two sectors, while $\langle \sigma v\rangle_{\rm co-SIMP}$ is the cross-section of the velocity averaged $s$-wave interaction. Because the resulting 21-cm signal is only sensitive to the product of these two terms, they are completely degenerate. To address this, in this work, we define a single, combined interaction parameter $\cint$, defined as
\begin{eqnarray}
\label{eq:cint}
    \cint \equiv \f\sqrt{\frac{M_{\chi}c^2}{0.1\, {\rm MeV}}}\,\sqrt{\frac{\left(0.5\, {\rm MeV}\right)^3}{\left(M_{\psi}c^2\right)^3}}\,\frac{\langle \sigma v\rangle_{\rm co-SIMP}}{1.5\times10^{-22} {\rm cm}^3/{\rm s}},
\end{eqnarray}
which serves to quantify the properties of the DM model. The scaling of each quantity in the above expression is motivated by observational results from the EDGES collaboration \citep{Paul:2023emc}. Physically $\f$ varies within $[0,2]$, while the factor $\frac{\langle \sigma v\rangle_{\rm co-SIMP}}{1.5\times10^{-22} {\rm cm}^3/{\rm s}}$ is allowed to vary within $[0.5,1.5]$. Consequently, $\cint$ spans the range $[0,3]$. $\cint\to 0$ marks the non-interacting condition between the dark and standard sectors. Thus, at $\cint\to0$, the co-SIMP model boils down to the standard CDM scenario.

By numerically solving the above set of equations, we can calculate the sky-averaged differential brightness temperature ($\deltat$) at any redshift during the dark ages. We depict the redshift evolution ($z\in[30-200]$) of $\deltat$ in Fig.~\ref{fig:global}. The figure illustrates how the value of $\cint \in [0-3]$ influences the global signal during dark ages, compared to the standard CDM scenario (black dashed line), showing higher values of $\cint$ increasing the strength of the signal. For the CDM case, the minimum of $\deltat$ is $-40.57$ mK, occurring at $z=85.6$. The minimum value decreases further with increasing $\cint$ and moves to high redshifts. For example, the minima of $\deltat$ are $-50.64$ mK, $-62.95$ mK, and $-78.31$ mK for $\cint=1,\,2~{\rm and}~3$, respectively, with the corresponding redshifts being $z=86.17,~86.75~{\rm and}~87.34$. Moreover, we found that for $z\lesssim50$, the dependence of $|\deltat|$ on $\cint$ starts to flip. $|\deltat|$ starts to decrease with increasing $\cint$, and even at some points it becomes smaller than the CDM scenario. This is because the DM is interacting with standard particles. This interaction reduces collisions (H-e, H-p, and H-H), thus reducing the collisional coupling. Consequently, the spin temperature couples with background radiation more rapidly than a non-interacting scenario.

\section{Signature on the 21-cm Power Spectrum}
\label{sec:darkage_ps}
While the global signal is a key observable, the power spectrum offers a much richer data set for probing the dark ages. The three-dimensional (3D) 21-cm power spectrum (PS), $\mathcal{P}_{21}(\textbf{k},\mu,z)$, quantifies the fluctuation of the 21-cm brightness temperature $\Delta T_{21} (\textbf{x}) \equiv T_{21}(\textbf{x}) - \deltat$ at a given $z$. $\mathcal{P}_{21}(\textbf{k},\mu,z)$ is defined via the Fourier transform as \citep{Munoz:2015eqa, Munoz:2016owz}
\begin{eqnarray}
\label{eq:def_P21}
    \langle \Tilde{\Delta T_{21}} (\textbf{k},z) \Tilde{\Delta T_{21}} (\textbf{k}',z) \rangle = \mathcal{P}_{21}(\textbf{k},\mu,z) (2\pi)^3 \delta_D(\textbf{k} - \textbf{k}').
\end{eqnarray}
Here, $\Tilde{\Delta T_{21}} (\textbf{k},z)$ is the Fourier transformation of $\Delta T_{21} (\textbf{x},z)$. $\mathcal{P}_{21}(\textbf{k},\mu,z)$ can be reduced to spherical (1D) 21-cm PS, $P_{21}(k,\mu,z)$, by integrating $\mathcal{P}_{21}(\textbf{k},\mu,z)$ over the solid angle in k-space. In the linear regime, the 1D 21-cm PS can be expressed analytically, including the anisotropy from peculiar velocities, as \citep{Munoz:2015eqa}
\begin{eqnarray}
\label{eq:p21}
    P_{21}(k,\mu,z)=\left(\mathcal{A}_{21} (z) + \deltat(z) \mu^2\right)^2 P_{\rm HI}(k,z),
\end{eqnarray}
where, $\mu\equiv \frac{k_{||}}{k}$ is the cosine of the angle between the line-of-sight ($k_{||}$) and the wavevector ($\textbf{k}$). Here, $P_{\rm HI}(k,z)$ represents the PS for the neutral hydrogen density fluctuation, which at the redshifts of interest, can be approximated as the baryon PS $P_{\rm b}(k,z)$. This is because the baryon contribution mostly arises from neutral hydrogen during the dark ages. We numerically calculate $P_{\rm b}(k,z)$ using a modified version of the publicly available Boltzmann solver code \texttt{CLASS}\footnote{\url{https://github.com/lesgourg/class_public}} \citep{2011arXiv1104.2932L,2011JCAP...07..034B}, as detailed in \cite{Paul:2023emc}. The coefficient $\mathcal{A}_{21}(z)$ represents the first-order derivative of the brightness temperature with respect to the baryon density fluctuation \textit{i.e.} $\mathcal{A}_{21}(z)\equiv \frac{\partial T_{21}(\textbf{k},z)}{\partial \delta_b}$ \citep{Munoz:2015eqa, Pillepich:2006fj}. For this term, we adopt the analytic form following \cite{Pillepich:2006fj}, which is discussed in detail in the Appendix~\ref{app:A_21_cal}. Throughout the study, we adopted the full-sky averaged \textit{i.e.} $\mu$-averaged 1D 21-cm PS which can be defined as~\footnote{This averaging procedure marginalizes over the anisotropic information inherent in the 21-cm signal due to redshift space distortions (RSD) and the light-cone effect. Although the $\mu$-averaging prevents the use of the Alcock-Paczynski effect to break degeneracies between astrophysical and cosmological parameters, $P_{21}(k,z)$ remains the standard metric for quantifying total signal variance and thermal noise sensitivity.}
\begin{eqnarray}
    \label{eq:mu_avg_PS}
    P_{21}(k,z)\equiv\frac{1}{2}\int_{-1}^{1}d\mu\, P_{21}(k,\mu,z).
\end{eqnarray}

\begin{figure*}[]
    \centering
    \begin{subfigure}{0.48\textwidth}
    \includegraphics[width=\textwidth]{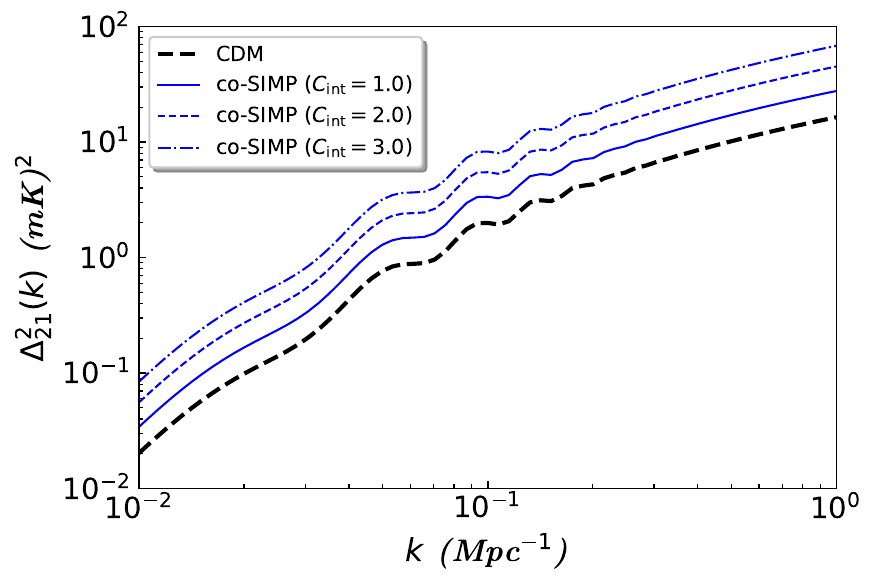}
    \caption{$z=60$.}
    \label{fig:p21_z60_cint_varry}
    \end{subfigure}\hfill
    \begin{subfigure}{0.48\textwidth}
    \includegraphics[width=\textwidth]{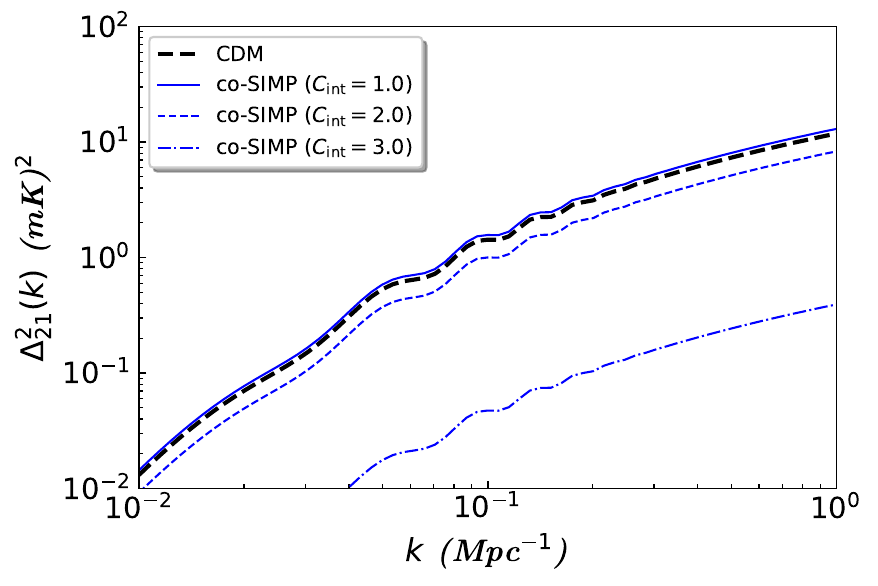}
    \caption{$z=40$.}
    \label{fig:p21_z40_cint_varry}   
    \end{subfigure}
    \caption{The 21-cm power spectra for the three different values of $\cint$ of co-SIMP model, compared with CDM. \textbf{Left panel} is for $z=60$, while \textbf{right panel} is for $z=40$.}
    \label{fig:PS_cint_varry}
\end{figure*}

The analytical expression of 21-cm PS (Eq.~\ref{eq:p21},\ref{eq:mu_avg_PS}) shows a strong dependence on the global brightness temperature, making the PS highly sensitive to $\cint$. In general, $\mu$-averaged 1D 21-cm PS is defined in such a way that dimension of k is absorbed. Thus, we defined $\Delta_{21}^2(k,z) \equiv k^3 P_{21}(k,z)/(2\pi^2)$ (in units of ${\rm mK}^2$). Fig.~\ref{fig:PS_cint_varry} shows the impacts of $\cint$ on the PS at $z=60$ and 40. As expected, the amplitude of the power spectrum increases with $k$, corresponding to a transition from large to small scales. The power spectra of the standard CDM and co-SIMP models exhibit nearly identical shapes, closely resembling that of the density power spectrum. However, the impact of $\cint$ on PS is non-trivial and evolves with $z$, directly mirroring the behaviour of the global signal as discussed in Sec.~\ref{sec:darkage_global}. At higher redshifts ($z\gtrsim 50$), where stronger interactions make the global signal stronger (Fig.~\ref{fig:global}), the 21-cm PS amplitude increases with $\cint$. This is illustrated for $z=60$ in Fig.~\ref{fig:p21_z60_cint_varry}. In contrast, the trend reverses at lower redshifts ($z\lesssim 50$). As discussed above, stronger interactions suppress the signal at these redshifts. Consequently, the amplitude of the 21-cm power spectrum decreases with $\cint$. This predicted change in behaviour is clearly demonstrated for $z=40$ in Fig.~\ref{fig:p21_z40_cint_varry}.

\begin{figure}
\centering
    \includegraphics[scale=0.58]{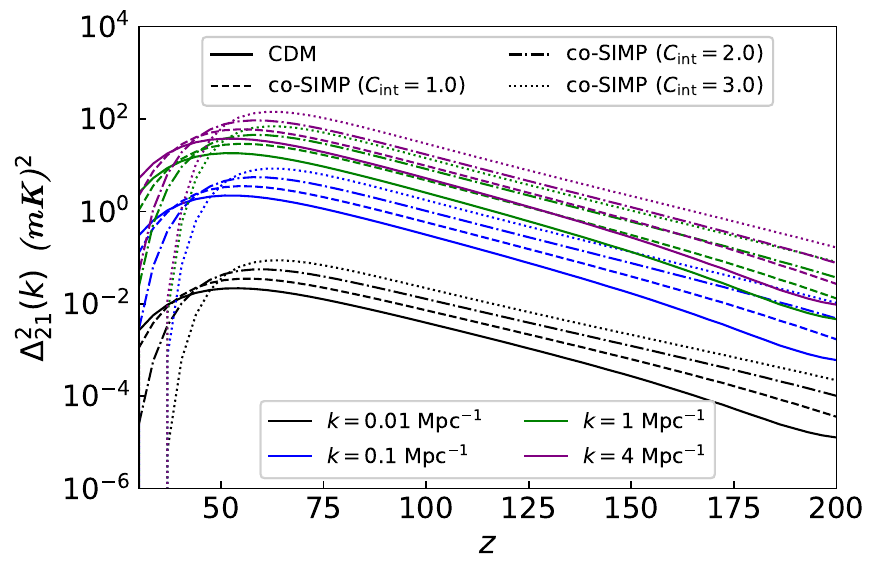}
    \caption{The redshift evolution of 21-cm power spectra for the co-SIMP models for three different values of $\cint$, compared with the CDM model. The power spectra are plotted at four values of $k$ (Mpc$^{-1}$). For $z\gtrsim 50$, the power spectrum increases with stronger DM interactions, whereas the trend reverses for $z<50$.}
    \label{fig:noise_spectra_zprofile}
\end{figure}

To understand in detail the redshift evolution of the 21-cm power spectra, we present its profile in Fig.~\ref{fig:noise_spectra_zprofile} at four representative wavenumbers $k=[0.01,0.1,1,4]$ (Mpc$^{-1}$). For each $k$, we compare the redshift profiles of the power spectrum for the standard CDM with the co-SIMP model (for $\cint=[1,\,2,\,3]$). For all models, the power spectrum rises with time to a peak and then declines. The initial rise is driven by the growth of fluctuations and the increase in the signal strength as the gas cools below the CMB temperature. The subsequent fall occurs at lower redshifts as the decreasing gas density weakens $x_{\rm c}$, reducing the signal. The peak in the power spectrum is a direct manifestation of the absorption trough in the global signal. The figure illustrates that as $\cint$ increases, this peak systematically shifts to higher redshifts, mirroring the behaviour of the global signal (Fig.~\ref{fig:global}).

\section{Prospects of detections}
\label{sec:prospects}
Noise from various sources significantly influences the observations and affects the accuracy of our predictions. A detailed understanding of the noise is crucial to identifying and mitigating its impacts. Statistical uncertainties and the specific experimental setup contribute to errors in the measurements of the global signal and PS.

\subsection{Global signal}
For the global signal, the dominant experimental challenge is overcoming thermal noise from bright astrophysical foregrounds. For a given observational bandwidth ($\Delta \nu$) and total integration time ($t_{\rm int}$), the thermal noise can be estimated as \citep{Shaver:1999gb}
\begin{eqnarray}
\label{eq:thermal_noise_t21}
    T_{\rm thermal}^{\rm N} \approx \frac{T_{\rm sys}}{\sqrt{\Delta\nu\; t_{\rm int}}}\,.
\end{eqnarray}
Here, we assume that the system temperature ($T_{\rm sys}$) is dominated by the sky brightness temperature, for which we adopt the form $T_{\rm sky} \approx 180\times\left(\frac{\nu}{180 {\rm MHz}}\right)^{-2.6}$\,K \citep{Furlanetto:2006jb}. Thus, thermal noise increases steeply with redshift as $T_{\rm thermal}^{\rm N}\propto z^{2.6}$. Given the presence of this noise, the expected significance of a detection at a specific $z$ can be quantified by the signal-to-noise ratio (SNR), which is calculated as
\begin{eqnarray}
    {\rm SNR}\Big|_{\rm global} \equiv \frac{\left|\deltat\right|}{T^{\rm N}_{\rm thermal}}.
\end{eqnarray}
Thus, increasing the integration time improves the detection prospect by reducing the noise spectrum (\textit{c.f.} Eq.~\eqref{eq:thermal_noise_t21}). Longer integration times can be achievable by employing a network (multiple copies) of global antennas. To perform the SNR analysis, we adopt an optimistic integration time of $t_{\rm int}=10,000$ hours.

\begin{figure}
    \centering
    \includegraphics[scale=0.6]{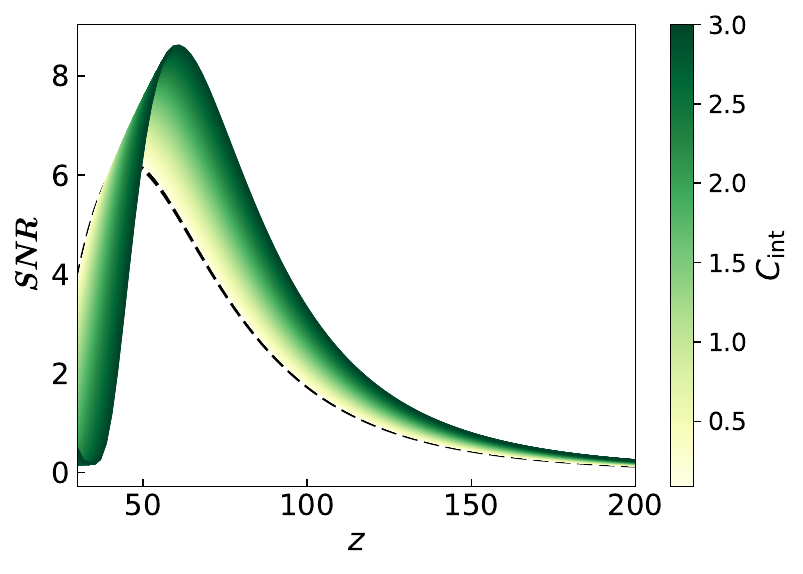}
    \caption{The SNR for the global 21-cm signal measurements as a function $z$, considering various $\cint$, for $t_{\rm int}=10,000$ hours and $\Delta \nu=1$\,MHz. The black dashed line represents the CDM scenario. Co-SIMP interaction enhances the SNR for $z\gtrsim 50$, however, the trend reverses for $z<50$.}
    \label{fig:snr_global}
\end{figure}


Fig.~\ref{fig:snr_global} shows the expected SNR for the global 21-cm signal measurements as a function of $z$, considering various $\cint$, for $t_{\rm int}=10,000$ hours and $\Delta \nu=1$\,MHz. The black dashed line corresponds to the standard CDM scenario. The co-SIMP interaction significantly affects the global signal strength (Fig.\ref{fig:global}). Although the standard CDM model reaches a maximum SNR of $13.96$ at $z= 45.59$, the co-SIMP interaction enhances this peak and shifts it to higher redshifts. For example, the co-SIMP scenario with $\cint=1$ achieves a maximum SNR of $15.73$ at $z= 49.99$, which increases further to $19.26$ at $z= 60.58$ for $\cint=3$. This enhancement demonstrates that a stronger co-SIMP interaction could make the global signal more likely to be detectable during the dark ages. 

To forecast the ability of future lunar and space-based global experiments to detect the signal from the dark ages, it is essential to account for contamination from astrophysical foregrounds, which are dominated by galactic synchrotron radiation. In our analysis, we adopt the formalism for calculating the significance with which we can measure the global signal presented in \cite{Mondal:2023bxb}. Our model is a combination of two terms. The first is a parameter $\beta$ times the expected signal, and the second term is a foreground contaminating component characterized by a power-law spectral shape, the same as the sky brightness temperature, \textit{i.e.} $A {\nu}^{-2.6}$. The amplitude of this power law, $A$, is treated as a free parameter\footnote{This is an optimistic benchmark for future space or lunar missions, where the absence of ionospheric distortion and radio frequency interference (RFI) should simply foreground removal compared to terrestrial experiments. As discussed in \cite{Park:2025phj}, the Dark Ages signal is less degenerate with foregrounds than the CD/EoR signal. While including additional parameters (e.g., freeing the spectral index) increases the degeneracy and reduces significance, the qualitative prospects for detecting the Dark Ages signal remain robust. For instance, moving from a 1-parameter to a 2-parameter model reduces the significance by a factor of $\sim 2$.} in the analysis, allowing us to marginalize over the uncertainty introduced by the removal of the smooth foreground component. The scaling parameter $\beta$ quantifies the statistical significance of the potential detection by modulating the amplitude of the predicted global signal. A value of $\beta=1$ corresponds to the full signal predicted by a given model, whereas $\beta=0$ represents the null hypothesis (\textit{i.e.}, no signal). This parametrization reformulates the problem as a statistical test of how well the data can distinguish the signal from zero, given the observational noise and the foreground model. While $C_{\rm int}$ is the physical parameter of the model, we utilize $\beta$ interpolation for the Fisher forecast to isolate the discriminatory power of the signal shape itself. In a Fisher analysis, using $C_{\rm int}$ directly can be problematic near $C_{\rm int}=0$ because the Fisher information matrix assumes the likelihood is locally Gaussian, but the $\Lambda$CDM model sits at the boundary of the $C_{\rm int}$ parameter space. By using $\beta$, we linearized the separation between the two models. Therefore, it provides a model-independent measure of how many standard deviations a co-SIMP model is from the $\Lambda$CDM model. This approach is a standard technique in 21-cm cosmology (e.g., \citep{Mondal:2023xjx, Mondal:2023bxb, Park:2025phj}) and is mathematically equivalent to a frequentist profile-likelihood ratio test between two fixed hypotheses. We acknowledge that $\beta$ is phenomenological, but at the points $\beta=0$ and $\beta=1$, it maps exactly to the physical models under consideration.

We perform a Fisher matrix analysis to forecast the measurement uncertainty $\delta \beta$ when measuring $\beta$. The statistical significance of detecting the signal is then given by $1/\delta \beta$, which represents the $Z$-score (the number of standard deviations $N\sigma$) by which the benchmark model ($\beta=1$) deviates from the null hypothesis ($\beta=0$). We adopt a one-tailed statistical test, as the model predicts a specific direction of deviation (i.e., we are testing the hypothesis $\beta > 0$). In this framework, the $1/\delta \beta$ value corresponds to the confidence level under the Gaussian assumption. For example, a significance of $1.645 \sigma$ corresponds to a 95\% confidence level (CL) for a one-tailed distribution. Throughout our analysis, all other cosmological parameters are fixed at their best-fit values, as mentioned above.

For our Fisher analysis, we assume a hypothetical experiment that observes the global signal at 40 redshift values spanning a frequency range corresponding to redshifts $30 \lesssim z \leq 200$, with a frequency resolution of $\Delta \nu=1$ MHz. We forecast the significance of the detection for three integration times: 1,000, 10,000 and 100,000 hours, which are achievable by employing a network of global antennas. The results of this forecast are tabulated in the upper panel of Table~\ref{tab:ERB}. The significance of detecting the standard CDM signal ($\cint=0$) is $3.939 \sigma$ for an integration time of 1,000 hours, consistent with the findings of \cite{Mondal:2023bxb}. The global signal would be distinguishable from zero for the co-SIMP model with even higher confidence. For example, for the same integration time of 1,000 hours, the detection significance increases to $4.304\sigma$, $5.234 \sigma$ and $6.161 \sigma$ for interaction strengths $\cint=1$, 2 and 3, respectively. For $t_{\rm int} = 10,000$\,hours and $100,000$\,hours, the detection significance increases by $\sqrt{10}$ times and 10 times, respectively.

Beyond merely detecting a signal, we next consider a more challenging detection scenario: forecasting how well an experiment could distinguish the co-SIMP models from the standard CDM signal. A statistically significant distinction would serve as robust evidence that the signal is different from the standard signal and must correspond to exotic physics. To perform this test, the model for our Fisher analysis is reformulated. In this case, the model is constructed as the term that accounts for the removal of the foreground plus the standard CDM signal plus $\beta$ times the difference between the predictions of the co-SIMP and the standard model:
\begin{equation}
    S_{\rm model}= A {\nu}^{-2.6}+ S_{\rm CDM} + \beta (S_{\rm co-SIMP} - S_{\rm CDM})\,.
\end{equation}
In this framework, the null hypothesis is now the standard CDM model (\textit{i.e.} $\beta = 0$). This analysis tests how significantly a non-zero value of $\beta$ can be measured, with $\beta = 1$ corresponding to full confirmation. The forecast results for this second scenario are tabulated in the lower panel of Table~\ref{tab:ERB}. For an integration time of 1,000 hours, the statistical detection significances are $1.595\sigma$, $4.149 \sigma$ and $6.572 \sigma$ for interaction strengths $\cint=1$, 2 and 3, respectively. For other integration times, the detection significance for all models improves as expected with $\sqrt{t_{\rm int}}$. This analysis reveals a noteworthy and seemingly counter-intuitive result if we notice closely. For $\cint=3$, the model is more easily distinguished from the standard CDM signal (at $6.572\sigma$) than from a complete absence of a signal (at $6.161 \sigma$). The reason for this outcome is discussed in the next paragraph.

\begin{figure}
    \centering
    \includegraphics[scale=0.54]{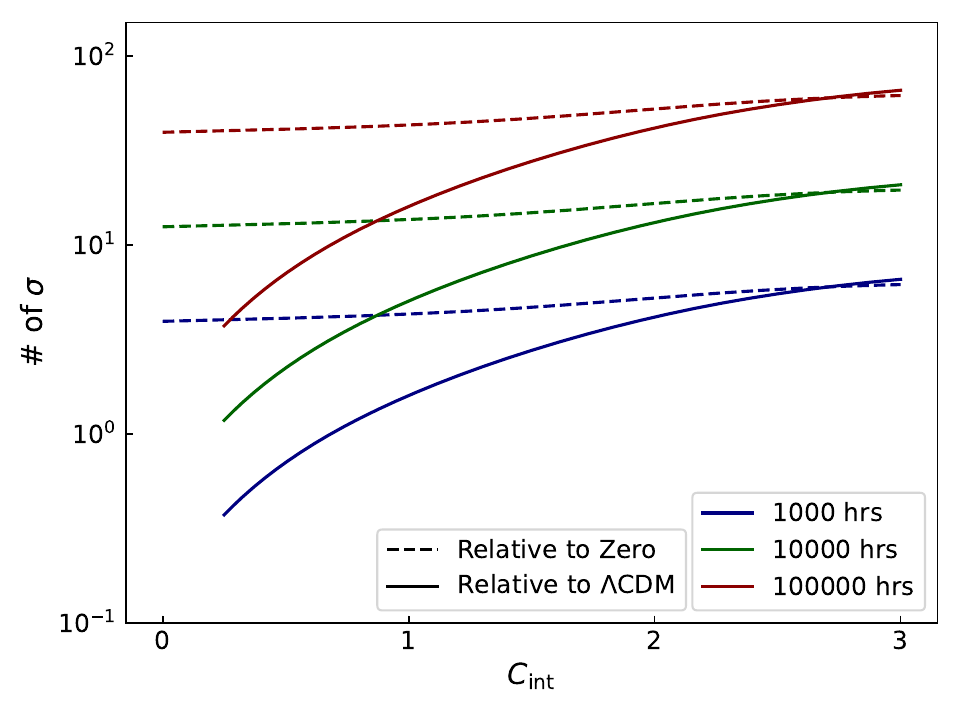}
    \caption{The significance (\# of $\sigma$) of the detection as a function of $\cint$ for the 21-cm global signal measurements for various integration times.}  
    \label{fig:global_sig}
\end{figure}

The statistical significance of detecting the global 21-cm signal for the co-SIMP model, quantified as the number of standard deviations ($\sigma$) by which it can be distinguished from zero or from the standard CDM signal, is shown in Fig.~\ref{fig:global_sig}. The figure shows this significance as a function of $\cint$ at three representative values of the integration times $t_{\rm int}=1,000$\,hours, 10,000\,hours and 100,000\,hours. As expected, the significance increases with $\cint$ and $t_{\rm int}$. For example, for the co-SIMP model, global signal can be detected at 5.23$\sigma$ significance relative to zero and distinguished from the standard CDM model at 4.15$\sigma$ for $\cint=2$ and 1,000\,hours integration. This indicates remarkable detection prospects even with modest observation times. As expected, it is easier to distinguish the co-SIMP signal from zero than from the standard CDM model for lower values of the interaction term ($\cint$). However, this trend reverses for $\cint \gtrsim 2.75$. The reason for this crossover lies in how the signal shapes differ in the most sensitive part of the redshift range $30<z<50$. As shown in Figs.~\ref{fig:global} and \ref{fig:snr_global}, a significant portion of the statistical sensitivity arises from lower redshifts ($30<z<50$) where the thermal noise is lowest. In this crucial redshift range, the co-SIMP signal for larger values of $\cint$ approaches zero faster than the standard CDM signal. Consequently, the absolute difference between the co-SIMP and CDM signal can become larger than the amplitude of the co-SIMP signal itself in this redshift range. Since the Fisher analysis integrates information across the entire redshift range, this large signal difference in that sensitive range makes it statistically easier to distinguish co-SIMP models with $\cint \gtrsim 2.75$ from CDM than from a complete absence of a signal.

\begin{table*}
\centering
\renewcommand{\arraystretch}{1.2}
\begin{minipage}{0.9\textwidth}
\centering
\begin{tabular*}{\textwidth}{c|c|c@{\extracolsep{\fill}}cc}
\hline \hline
\multirow{2}{*}{\textit{Comparison}} & \multirow{2}{*}{$\cint$} & \multicolumn{3}{@{}c@{}}{\textit{Integration Time}} \\
\cline{3-5}
{} & {} & 1,000 {hrs} & 10,000 {hrs} & 100,000 {hrs} \\
\hline

\multirow{4}{*}{Relative to zero}
  & 0 (CDM) & 3.939	& 12.457 &	39.392 \\
  & 1.0	& 4.304	& 13.610 &	43.039 \\
  & 2.0	& 5.234	& 16.550 &	52.336 \\
  & 3.0	& 6.161	& 19.482 &	61.606 \\

\hline

\multirow{3}{*}{Relative to standard}
  & 1.0	& 1.595 &	5.044 &	15.949 \\
  & 2.0	& 4.149	& 13.121 &	41.492 \\
  & 3.0	& 6.572 & 20.783 & 	65.722\\

\hline \hline
\end{tabular*}
\end{minipage}
\caption{The significance (\# of $\sigma$) of detecting the co-SIMP global signal relative to zero signal and standard CDM signal ($\cint=0$), for various values of $\cint$ and integration times.}
\label{tab:ERB}
\end{table*}

\subsection{Power spectrum}
Although the 21-cm PS is a much richer dataset than the global signal, the presence of unavoidable noise makes its measurement challenging. The total error in a power spectrum measurement has two primary contributions: thermal noise and cosmic variance (CV).

The CV is an intrinsic, sample-level uncertainty arising from the fundamental limitation that we can observe only a single Universe. Assuming the 21-cm signal being a Gaussian random field \citep{Mondal2015}, the noise contribution from CV to a measurement of $P_{21}(k,\nu)$ at a specific wavenumber $k$ and frequency $\nu$, can be expressed as \citep{Mondal:2015oga}: 
\begin{eqnarray}
\label{eq:CV_noise}
    P^{\rm N}_{\rm CV}(k,\nu) \equiv \frac{2\pi P_{21}(k,\nu)}{\sqrt{V(\nu)k^3\Delta(\ln k)}}\,,
\end{eqnarray}
where $V(\nu)$ represents the survey volume within a frequency bin centred at $\nu$ with a bandwidth of $\Delta \nu$. This volume is given by $V(\nu)=\Omega_{\!_{\rm FOV}}r_{\nu}^2\Delta r_{\Delta \nu}$, where $r_{\nu}$ is the comoving distance to that frequency bin and $\Delta r_{\Delta \nu}$ is the comoving depth corresponding to the bandwidth. The field of view, $\Omega_{\!_{\rm FOV}}$, is related to the wavelength of observation ($\lambda_{\rm obs}$) and the effective area of an instrument $A_{\rm eff}$ as $\Omega_{\!_{\rm FOV}} = \lambda_{\rm obs}^2/A_{\rm eff}$. During the dark ages, the field of view can be a significant fraction of the sky. Therefore, following \cite{Mondal:2023xjx}, we set a maximum available solid angle of $2\pi$ steradians (half the sky) for our analysis. Eq.~\eqref{eq:CV_noise} reveals that that for a constant $\Delta(\ln k)$, the CV scales as $k^{-3/2}$. This indicates that CV is the dominant source of uncertainty on large scales (\textit{i.e.}, low $k$).

The second major source of noise is thermal noise. The contribution of this noise to the measurement of the power spectrum can be expressed as \citep{2013ExA....36..235M}
\begin{eqnarray}
\label{eq:thermal_noise}
    P^{\rm N}_{\rm therm}(k,\nu) \equiv \sqrt{\frac{4\, k^3\,V(\nu)}{\pi^2\,\Delta(\ln k)}}\, \frac{T_{\rm sys}^2}{\Delta\nu\, t_{\rm int}}\, \frac{A_{\rm core}}{N_{\rm st}^2 \, A_{\rm eff}}\,,
\end{eqnarray}
where $A_{\rm core}$ is the core area and $N_{\rm st}$ is the total number of stations in the telescope array. Following \cite{Mondal:2023xjx}, we assume that $A_{\rm eff}=25\,{\rm m}^2$ and the total collection area is equal to the core area, such that $A_{\rm coll} = A_{\rm core} = N_{\rm st} \times A_{\rm eff}$. With this setup, the overall dependence of the thermal noise associated with power spectrum scales as $P^{\rm N}_{\rm therm} \propto 1/(N_{\rm st} \, t_{\rm int})$. Keeping this relation fixed, we vary $A_{\rm coll}$ and $t_{\rm int}$. From Eq.~\ref{eq:thermal_noise}, we also see that the thermal noise scales as $k^{3/2}$. This means that thermal noise is the dominant source on small scales (\textit{i.e.}, at high $k$).

\begin{table}
\begin{center}
\renewcommand{\arraystretch}{1.2}
\begin{minipage}{8.5cm}
\begin{tabular*}
{\textwidth}{l|c@{\extracolsep{\fill}}cc}
\hline
\hline

& \multicolumn{3}{@{}c@{}}{\textit{Configuration}}\\
\cline{2-4}
{} & G & A & B\\

\hline

$A_{\rm coll}$ [km$^2$] & 5 & 10 & 10 \\
$t_{\rm int}$ [hrs] & 1,000 & 1,000 & 10,000 \\

\hline
\hline
\end{tabular*}
\end{minipage}
\caption{The 21-cm power spectrum observational configurations in terms of the collecting area $A_{\rm coll}$ and integration time $t_{\rm int}$.}
\label{tab:conf}
\end{center}
\end{table}

The resultant total noise in the measurement of the power spectrum is estimated by combining the cosmic variance and thermal noise contributions as $P_{21}^{\rm N} (k,z)\equiv \sqrt{(P_{\rm CV}^{\rm N}(k,z))^2+(P_{\rm therm}^{\rm N} (k,z))^2}$, with $\nu=1420\,{\rm MHz}/(1+z)$. The SNR for the power spectrum can be defined as 
\begin{eqnarray}
{\rm SNR}\Big|_{\rm PS} \equiv \frac{P_{21}(k,z)}{P_{21}^{\rm N} (k,z)}\,.
\end{eqnarray}
This is a function of $k$, $z$ and the interaction parameter $\cint$. To illustrate the observational prospects, we considered three observational configurations, specified in terms of $t_{\rm int}$ and $A_{\rm coll}$, as tabulated in Table~\ref{tab:conf}. Here, the configuration G is designed to roughly match the statistical power of the global signal case for $t_{\rm int} = 1,000$\,hours \citep{Mondal:2023xjx}. Throughout this analysis, we assume $\Delta(\ln k)=0.5$ and $\Delta \nu=5$\,MHz.

\begin{figure*}[]
    \centering
    \includegraphics[scale=0.45]{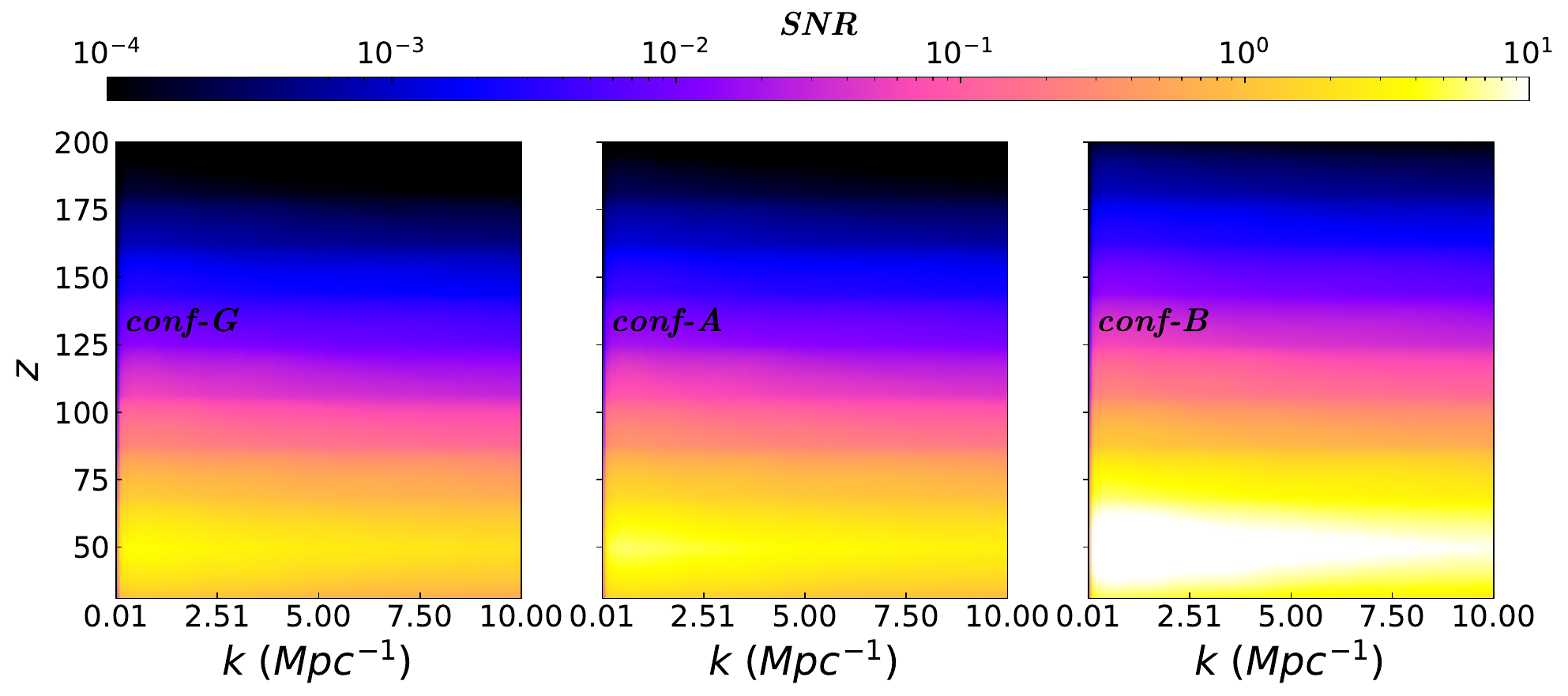}
    \caption{Illustration of SNR in $k$-$z$ plane for the co-SIMP model for $\cint=1.0$. The panels corresponding to the experimental configurations \textit{G}, \textit{A} and \textit{B}, demonstrate that the detectability of the 21-cm signal is maximised for $k=0.46$ Mpc$^{-1}$ and $z=49.78$. Among these, configuration \textit{B} provides the more optimistic sensitivity for probing the distinctive features of co-SIMP DM models through the 21-cm signal.}
    \label{fig:snr_fixed_cint}
\end{figure*}

Fig.~\ref{fig:snr_fixed_cint} shows the variation of the SNR in the $k-z$ plane for $\cint=1$. For all three configurations, the SNR peaks at $k=0.46$ Mpc$^{-1}$ and $z=49.78$. This behaviour arises because the mid-$k$ regime is less affected by the dominance of cosmic variance at low-$k$ and thermal noise at high-$k$. Furthermore, the SNR peaks around $z=50$, corresponding to the epoch where the amplitude of the power spectrum itself is maximal (see Fig.~\ref{fig:noise_spectra_zprofile}). For the minimal experimental setup (configuration G), the maximum SNR achieved is $4.06$ for $\cint=1.0$, which increases to $5.82$ and further to $18.405$ under improved experimental setups, configuration A and B, respectively.

\begin{figure*}[]
    \centering
    \includegraphics[scale=0.45]{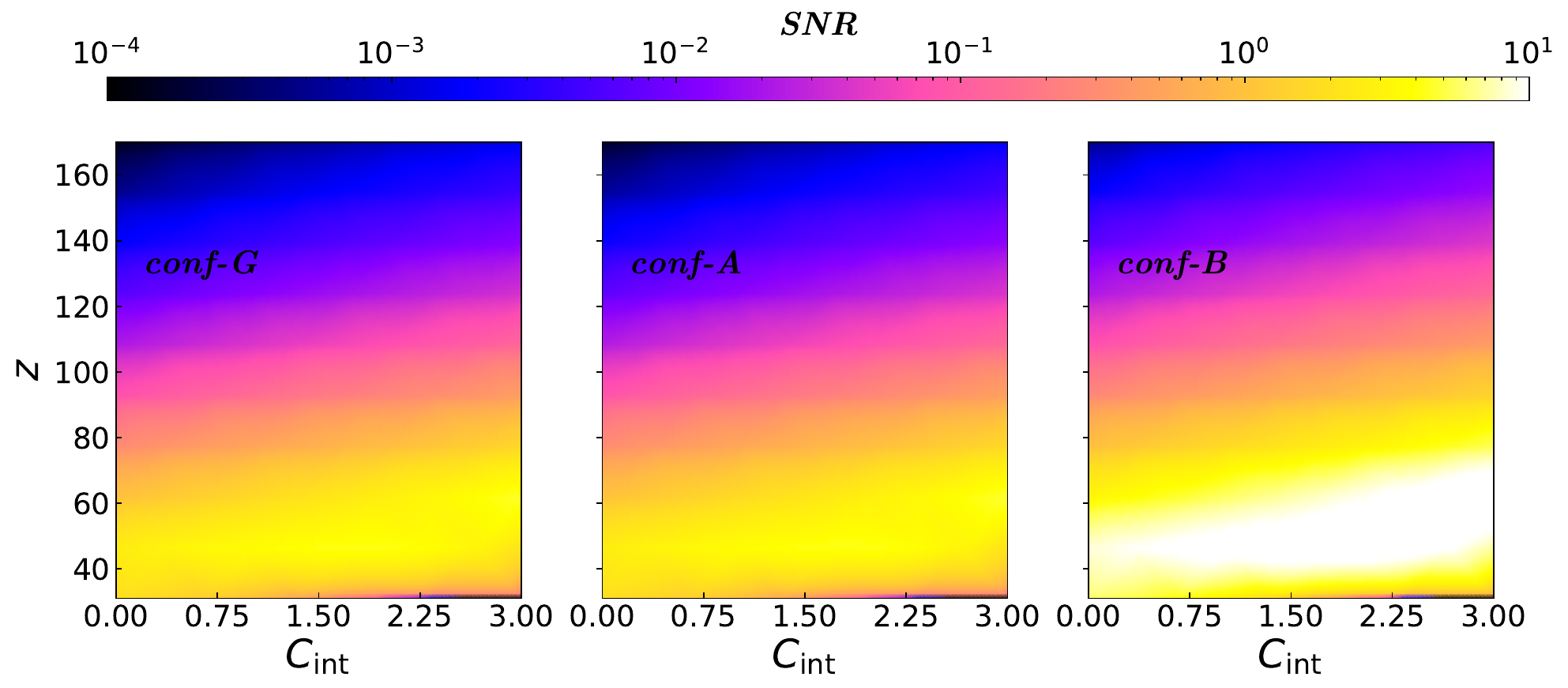}
    \caption{Illustration of SNR in $\cint$-$z$ plane for $k=0.46$ Mpc$^{-1}$, considering the three configurations. In all the configurations, the SNR attains its maximum at $z=61.89$ and $C_{\rm int}=3.0$. Consistent with Fig.~\ref{fig:snr_fixed_cint}, configuration \textit{B} represents the more optimistic experimental setup for probing the co-SIMP DM models via the detection of 21-cm signal.}
    \label{fig:snr_fixed_k}
\end{figure*}

\begin{figure*}[]
    \centering
    \includegraphics[scale=0.45]{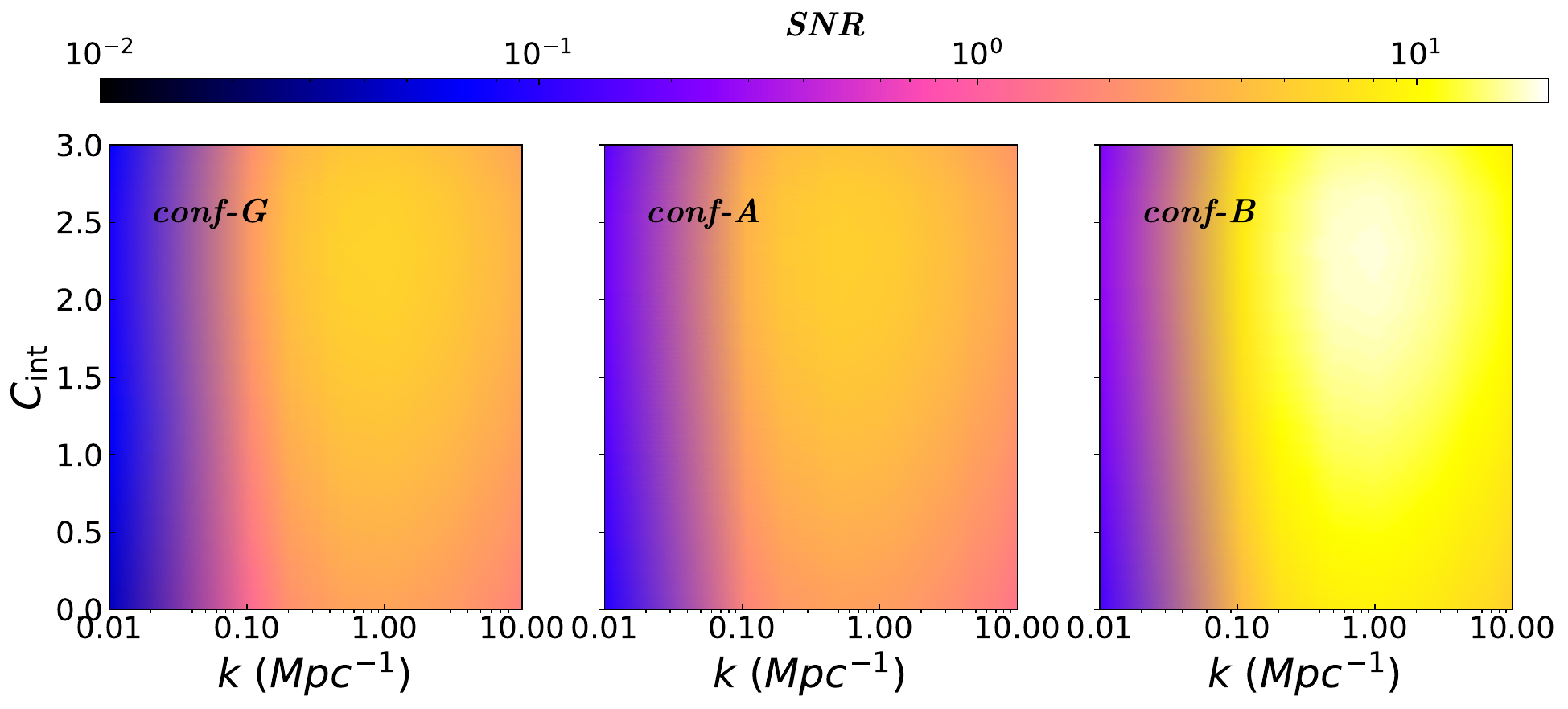}
    \caption{Illustration of SNR in $k$-$\cint$ plane, considering the three configurations at $z=50$. SNR attains its maximum for $k=1.05$ Mpc$^{-1}$ and $c_{\rm int}=2.5$, in all the cases. Consistent with Fig.~\ref{fig:snr_fixed_cint}, among the three configurations, configuration \textit{B} exhibits the highest sensitivity, making it the most optimistic setup for probing the characteristic features of the co-SIMP model through the detection of the 21-cm signal.}
    \label{fig:snr_fixed_z}
\end{figure*}

To explore this dependence more comprehensively, we set the wavenumber at its optimal value of $k=0.46$ Mpc$^{-1}$ and depict the variation of SNR in the $\cint-z$ plane in Fig.~\ref{fig:snr_fixed_k}. Here, the overall maximum occurs at $z=61.89$ with $\cint=3.0$. This behaviour is physically driven by the fact that for high redshifts ($z\gtrsim 50$), the power spectrum amplitude increases with $\cint$, whereas the trend begins to reverse for $z\lesssim50$ (as discussed in Sec.~\ref{sec:darkage_ps}). For configuration G, the maximum SNR found in this plane is $4.49$. This value increases to $4.61$ and $14.58$ for configurations A and B, respectively. Furthermore, the same is explored in the $k-\cint$ plane at $z=50$, as shown in Fig.~\ref{fig:snr_fixed_z}. The figure illustrates that the maximum SNR reaches $5.44$ for configuration G, which increases to $5.73$ and $18.12$ for configurations A and B, respectively. Thus, a broader exploration of the parameter space shows that the maximum SNR for Configuration~G is $5.44$, attained at $z = 49.78$, $k = 1.0\,\mathrm{Mpc}^{-1}$, and $\cint = 2.3$. For Configurations~A and~B, the corresponding maximum SNR values increase to $5.73$ and $18.12$, respectively.

\begin{table*}
\centering
\begin{minipage}{0.9\textwidth}
\centering
\renewcommand{\arraystretch}{1.2}
\begin{tabular*}{\textwidth}{c|c|c@{\extracolsep{\fill}}cc}
\hline \hline
 \multirow{2}{*}{\textit{Comparison}} & \multirow{2}{*}{$\cint$} & \multicolumn{3}{c}{\textit{Configuration}} \\
\cline{3-5}
 & & G & A & B \\
\hline

\multirow{4}{*}{Relative to zero}
 & 0 (CDM)	& 5.446	& 10.893 & 108.822 \\
 & 1.0 & 4.632 & 9.264 & 92.613 \\
 & 2.0 & 3.064	& 6.127	& 61.268 \\
 & 3.0 & 1.914 & 3.829 &	38.285 \\
\hline

\multirow{3}{*}{Relative to standard}
 & 1.0 & 1.778 &	3.556 &	35.552 \\
 & 2.0 & 4.292 & 8.584 & 85.842 \\
 & 3.0 & 5.336 &	10.671 & 106.714 \\

\hline \hline
\end{tabular*}
\end{minipage}
\caption{The significance (\# of $\sigma$) of detecting the co-SIMP power spectrum relative to zero or the standard CDM signal.}
\label{tab:ERBps}
\end{table*}

\begin{figure}
    \centering
    \includegraphics[scale=0.54]{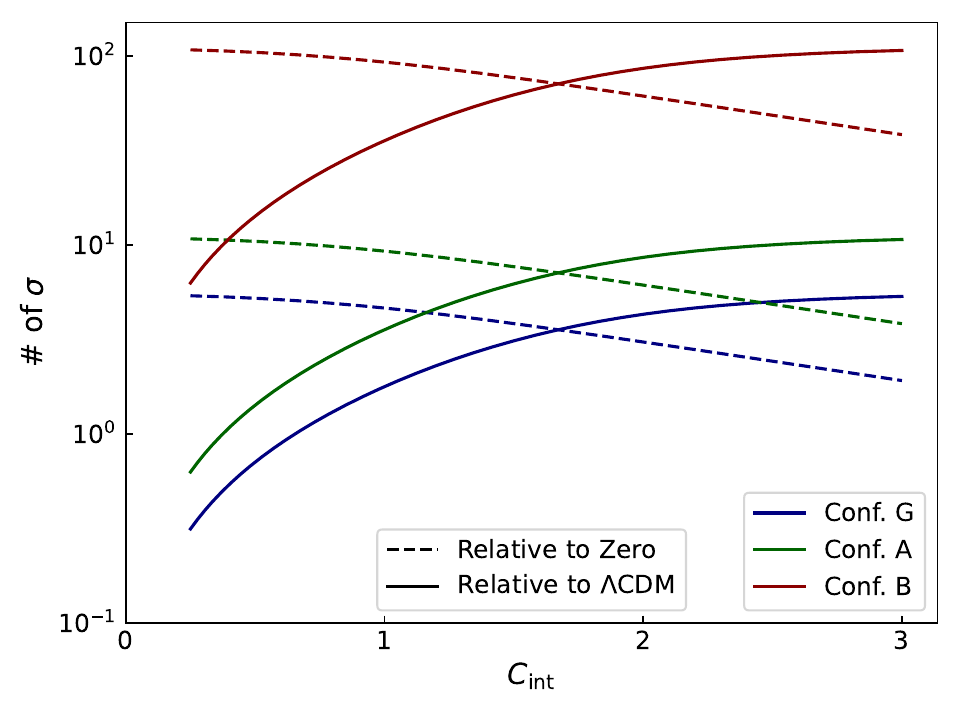}
    \caption{The significance (\# of $\sigma$) of the detection as a function of $\cint$ for the 21-cm power spectra measurements for various configurations.}  
    \label{fig:power_spectra}
\end{figure}

Furthermore, we estimate the detection significance for measuring the spherically averaged 21-cm power spectrum, considering its evolution across different redshifts and scales. For this calculation, the Fisher analysis includes both CV noise (Eq.~\ref{eq:CV_noise}), which is important on large scales, and instrumental thermal noise (Eq.~\ref{eq:thermal_noise}). The analysis is performed using eight redshift bins that span the frequency range $8.305 \leq \nu \leq 43.305$ MHz (with $\Delta \nu=5$ MHz) and 11 logarithmically spaced wavenumber bins from $k= 0.01$ to 1.484 ${\rm Mpc}^{-1}$ (with bin width $\Delta (\ln{k})=0.5$). As with the global signal, we evaluate the statistical significance of detecting the power spectrum in two scenarios: (i) distinguishing the signal from a null hypothesis (zero signal), dominated by foregrounds and noise, and (ii) distinguishing the co-SIMP signal from the standard $\Lambda$CDM signal. 

For the minimal configuration G, representing a modest interferometric setup, the standard $\Lambda$CDM ($\cint=0$) signal can be detected relative to zero with $5.446 \sigma$ significance. For the same configuration, the co-SIMP signal with a small interaction strength $\cint=1.0$ can be detected relative to zero at a significance of $4.632 \sigma$. However, this significance decreases with increasing interaction strength, falling to $1.914 \sigma$ for $\cint =3.0$. This counter-intuitive trend is driven by two factors: suppression of the 21-cm power spectrum at a relatively low redshift ($z \leq 50$) where thermal noise is low, combined with a steep increase in thermal noise at higher redshifts where the signal for strong interactions is largest.

Although configurations with larger collecting areas considerably improve the detection prospects, this trend of decreasing significance with increasing $\cint$ remains. For example, configuration A achieves a $9.264 \sigma$ significance at $\cint=1.0$ and $3.829\sigma$ at $\cint =3.0$. Configuration B enables detection at $92.613 \sigma$ and $38.285 \sigma$, respectively. The details are presented in Table~\ref{tab:ERBps}.

In the second scenario, where we compare the co-SIMP power spectrum with the standard $\Lambda$CDM model, the detection significance increases monotonically with interaction strength. This is because a larger $\cint$ creates a greater deviation from the standard model, making the two signals easier to distinguish. In configuration G, the statistical significance increases from $1.778\sigma$ for $\cint = 1.0$ to $5.336\sigma$ for $\cint = 3.0$. The trend becomes even more pronounced with improved instrumental sensitivity. In configuration B, the significance increases from $35.552\sigma$ to $106.714\sigma$ over the same range of interaction strengths.

Fig.~\ref{fig:power_spectra} shows the significance (\# of $\sigma$) of the detection as a function of $\cint$ for the 21-cm power spectra measurements. As discussed above, the opposing trends of these two scenarios lead to a crossover point at $\cint \sim 1.75$. For smaller interaction strengths ($\cint \lesssim 1.75$), the co-SIMP power spectrum is statistically more distinguishable from the zero signal than the standard $\Lambda$CDM signal. However, beyond this crossover point, this reverses. 

\section{Conclusion}
\label{sec:conclusion}
The cosmic dark ages represent a pristine cosmological time period, and the 21-cm signal from this epoch is a potent probe with the capacity to constrain the microphysical properties of DM. Indeed, a growing body of research on non-standard scenarios suggests that exotic DM models could be more identifiable than the standard CDM signal. In this study, we explore the cosmological signatures of the co-SIMP DM model, focusing on its imprint on both the global 21-cm signal and its power spectrum. We introduce a combined dimensionless parameter, $\cint$, incorporating the masses of DM and SM particles, interaction cross-section between them, and the amount of heat exchange between the two sectors. $\cint=0$ represents the standard $\Lambda$CDM scenario. By parametrizing the combined parameter, we have forecasted the extent to which upcoming dark ages experiments can probe co-SIMP model and distinguish it from the standard CDM model.

Our analysis shows that the co-SIMP interaction has two primary imprints on the 21-cm signal. First, it enhances the amplitude of the global signal absorption trough and shifts its peak towards the higher redshifts. For example, while the CDM global signal peaks at $-40.57$ mK at $z=85.6$, the co-SIMP model with $\cint=1.0$ reaches a deeper $-50.64$ mK at $z=86.17$. Second, this enhancement is strongly dependent on redshift. At high redshifts ($z\gtrsim 50$), the interaction amplifies the signal, while at lower redshifts ($z\lesssim 50$), it leads to suppression due to the modified collisional coupling in the IGM (see Fig.~\ref{fig:global}). A similar redshift dependence is observed in the 21-cm power spectra. The power spectrum is enhanced with $\cint$ for $z\gtrsim50$, while for $z\lesssim50$ it leads to a suppression (see Figs.~\ref{fig:p21_z60_cint_varry} and \ref{fig:p21_z40_cint_varry}).

To properly assess the detectability of these features, we evaluate the SNR for various experimental setups. For $10,000$ hours of integration, the global signal achieves a SNR of $13.96$ at $z=45.59$ for the standard CDM model. Co-SIMP interaction significantly enhances the SNR, with its peak shifting toward the high redshifts. For example, the maximum SNR for $\cint=1.0$ is $15.73$ at $z=49.99$, which increases to $19.26$ for $\cint=3.0$, which occurs at $z=60.58$.

For power spectra, we first investigated the ranges of $k$ and $z$ at which the SNR achieves its maximum for a given interaction strength $\cint$. For example, under a minimal interferometric setup (configuration G), the co-SIMP model with $\cint=1.0$ achieves a maximum SNR of $4.06$ at $k=0.46$ Mpc$^{-1}$ and $z=49.78$. The value improves to $5.82$ for configuration A and $18.405$ for configuration B. 
Our broader exploration of the parameter space reveals that the overall maximum SNR for Configuration G is 5.44, occurring at $z=49.78$, $k=1.0$ Mpc$^{-1}$ and $\cint=2.3$. The maximum SNR value increases to $5.73$ and $18.12$ for configurations A and B, respectively.

Building on this, we performed the Fisher forecast analysis to quantify the prospect of detecting the signal in future experiments. For the global signal, an integration of $1,000$ hours can distinguish the co-SIMP model with $\cint =1.0$ from a null signal at $4.304\sigma$ and the standard CDM signal at $1.595\sigma$. The co-SIMP model with the strongest intersection of $\cint=3.0$ can be distinguished from the zero signal with $6.161\sigma$ and the standard CDM signal with $6.572\sigma$. The prospect increases significantly as the square root of the integration time, which is achievable with a network (multiple copies) of global antennas.

Given the richness of the 21-cm power spectrum dataset, we performed a similar Fisher forecast analysis for the interferometric measurement. Our study reveals two distinct opposing trends. When testing against a null signal, the significance decreases with interaction strength $\cint$. For configuration G, the model with $\cint=1.0$ can be detected at $4.632\sigma$, while $\cint=3.0$ is only detectable at $1.914\sigma$. In contrast, when distinguishing the co-SIMP signal from the standard CDM, the significance increases monotonically with $\cint$. For configuration G, the significance increases from 1.778$\sigma$ for $\cint=1.0$ to 5.336$\sigma$ for $\cint=3.0$. With an optimistic experimental setup such as configuration B, this distinction can be made with an extraordinary significance of 106.714$\sigma$.

In summary, our analysis demonstrates that the co-SIMP model leaves clear and distinguishable imprints on both the global 21-cm signal and its power spectrum during the dark ages. These features can be robustly detected with upcoming experiments, especially in configurations with extended integration times and larger antenna arrays. This study highlights the potential of 21-cm cosmology to probe the fundamental properties of dark matter and constrain particle physics beyond the standard CDM paradigm.

\section*{Acknowledgements}
The Authors thank Supratik Pal for fruitful discussions and insights. The Authors acknowledge the computational facilities of the Technology Innovation Hub, Indian Statistical Institute (ISI), Kolkata and the publicly available Boltzmann solver code \href{https://github.com/lesgourg/class_public}{\texttt{CLASS}}. The authors also thank the anonymous referee for valuable comments and insights that helped improve the manuscript. DP thanks ISI, Kolkata for financial support through Senior Research Fellowship. SP thanks CSIR for financial support through Senior Research Fellowship (File no. 09/093(0195)/2020-EMR-I). ADB acknowledges financial support from DST, India, under grant number IFA20-PH250 (INSPIRE Faculty Award). RM is supported by the NITC FRG Seed Grant (NITC/PRJ/PHY/2024-25/FRG/12).

\section*{Data availability}
The numerical calculations performed in the study is carried out using the python library \texttt{numpy} (\url{https://numpy.org/}), \texttt{matplotlib} (\url{https://matplotlib.org/}) and \texttt{scipy} (\url{https://scipy.org/}), and the publicly available Boltzmann solver \texttt{CLASS} (\url{https://github.com/lesgourg/class_public}). The data and the codes are available on request from the corresponding author.

\bibliographystyle{mnras}
\bibliography{mybib}

\appendix
\section{Estimation of total electron fraction}
\label{app:xe}
In this appendix, we will illustrate the estimation of the total ionization fraction of electrons ($x_{\rm e}$). The contributions of $x_{\rm e}$ arise from the ionization fraction of HI and HeI. Thus, $x_{\rm e}$ is considered as the sum of the proton fraction ($x_{\rm p}$) and the singly ionized helium fraction ($x_{\rm \!_{He}}$) \textit{i.e.} $x_{\rm e}=x_{\rm p} + x_{\rm \!_{He}}$. The redshift evolution of these two contributions to the ionization fractions can be estimated as \citep{Seager:1999bc}

\begin{eqnarray}
\label{eq:xp_evol}
    {dx_{\rm p}\over dz} &= \left(x_{\rm e}x_{\rm p} n_{\rm \!_{H}} 
    \alpha_{\rm \!_{H}}
    - \beta_{\rm \!_{H}} (1-x_{\rm p})
    {\rm e}^{-h_{\!_{P}} \nu_{\rm \!_{H2s}}/k_{\!_{B}} \tk}\right) \\
    &\times\quad{\left(1 + K_{\rm \!_{H}} \Lambda_{\rm \!_{H}} n_{\rm \!_{H}}(1-x_{\rm p})\right)
    \over H(z)(1+z)\left(1+K_{\rm \!_{H}} (\Lambda_{\rm \!_{H}} + \beta_{\rm \!_{H}})
     n_{\rm \!_{H}} (1-x_{\rm p}) \right)},\nonumber \\
     {dx_{\rm \!_{He}}\over dz} &=
   \left(x_{\rm \!_{He}}x_{\rm e} n_{\rm \!_{H}} \alpha_{\rm \!_{He}}
   - \beta_{\rm \!_{He}} (f_{\rm \!_{He}}-x_{\rm \!_{He}})
   {\rm e}^{-h_{\!_{P}} \nu_{\rm \!_{He2^1s}}/k_{\!_{B}} \tk}\right)\label{eq:xHe_evol}\\
   &\times {\left(1 + K_{\rm \!_{He}} \Lambda_{\rm \!_{\rm He}} n_{\rm \!_{H}}
  (f_{\rm \!_{He}}-x_{\rm \!_{He}}){\rm e}^{-h_{\!_{P}} \nu_{\rm \!_{ps}}/k_{\!_{B}} \tk})\right)
  \over H(z)(1+z)\left(1+K_{\rm \!_{He}}
  (\Lambda_{\rm \!_{He}} + \beta_{\rm \!_{He}}) n_{\rm \!_{H}} (f_{\rm \!_{He}}-x_{\rm \!_{He}})
  {\rm e}^{-h_{\!_{P}} \nu_{\rm \!_{ps}}/k_{\!_{B}} \tk}\right)}.\nonumber
\end{eqnarray}

In the above equations, the case B recombination coefficient, $\alpha$, has a different fitting formula for HI ($\alpha_{\rm H}$) and HeI ($\alpha_{\rm He}$). For HI, the fitting form has the following form \citep{1994MNRAS.268..109H,1991A&A...251..680P}:
\begin{equation}
\label{eq:alpha_H}
    \alpha_{\rm H} \equiv F\times 10^{-19}\frac{at^{b}}{1 + ct^{d}} \,\,  \mathrm{m^{3}s^{-1}},
\end{equation}
with, $a=4.309$, $b=-0.6166$, $c=0.6703$, $d=0.5300$, $F=1.14$ and $t\equiv \tk/10^{4}\,$K. The recombination coefficient for HeI has the following form~(\cite{hummer1998recombination})
\begin{equation}
\label{eq:alpha_He}
    \alpha_{\rm He} \equiv q\left[\sqrt{\tk\over T_2}\left(1+\sqrt{\tk \over T_2}\right)^{1-p}
    \left(1+\sqrt{\tk\over T_1}\right)^{1+p}\right]^{-1}\,
     \mathrm{m^{3}s^{-1}},
\end{equation}
with, $q=10^{-16.744}$, $p=0.711$, $T_1=10^{5.114}\,$K, and $T_2=3\,$K. The photoionization coefficient ($\beta$) in Eqs.~\eqref{eq:xp_evol} and \eqref{eq:xHe_evol} (with subscripts retaining the standard meaning), can be related to $\alpha$ as \citep{Seager:1999bc}
\begin{eqnarray}
\label{eq:beta}
    \beta \equiv \alpha\, \left(2\pi m_e k_{\!_{B}} \tk/h_P^2\right)^{3/2}\exp{\left(\frac{-h_P\,\nu_{2s}}{k_{\!_{B}} \tk}\right)},
\end{eqnarray}
where, $m_e$ is the electron mass. The redshifting of HI Ly-$\alpha$ photons and HeI $2^1 p- 1^1 s$ photons can be expressed as $K_{\rm H}\equiv \frac{\lambda^3_{\rm H}}{8\pi\,H(z)}$ and $K_{\rm He}\equiv \frac{\lambda^3_{\rm He}}{8\pi\,H(z)}$ \citep{Seager:1999bc}, respectively. The specific values of the parameters are tabulated in Table~\ref{tab:atomic_data}.

\begin{table}
    \centering
    \begin{tabular}{|c|c|c|}
        \hline
        Quantities & Physical meaning & Values\\
        \hline
        \hline
        $\lambda_{\rm H2p}$ & H Ly-$\alpha$ wavelength & $121.5682$ nm\\
        $\lambda_{\rm He}$ & He $2^1p-1^1s$ wavelength & $58.4334$ nm\\
        $\nu_{\rm H2s}$ & H $2s-1p$ frequency & $3.29\times 10^{15}$ Hz\\
        $\nu_{\rm He2^1s}$ & He $2^1s-1^1s$ frequency & $5\times 10^{15}$ Hz\\
        $\Lambda_{\rm H}$ & H $2s-1s$ two photon rate & $8.22458$ s$^{-1}$\\
        $\Lambda_{\rm He}$ & He $2^1s-1^1s$ two photon rate & $51.3$ s$^{-1}$\\
        \hline
    \end{tabular}
    \caption{Parameter values and their physical interpretation used in the analysis.}
    \label{tab:atomic_data}
\end{table}

\section{Calculation of \texorpdfstring{$\mathbf{\mathcal{A}_{21}}$}{A_21}}
\label{app:A_21_cal}
We have adopted the analysis of \cite{Pillepich:2006fj} to obtain the analytical expression for $\mathcal{A}_{21}(z)$. As mentioned in the main text, $\mathcal{A}_{21}(z)$ is the first order partial derivative of $T_{21}(\textbf{x},z)$ with respect to baryon over-density. In this appendix, we only focus on estimating the analytic expression for $A_{21}(z)$. $\TBRIGHT(\textbf{x},z)$ can be expressed up to the second order spacial fluctuation, in a general scenario, as
\begin{equation}
\TBRIGHT(\textbf{x},z) \equiv \TBRIGHT^{(0)}(z)+ \delta T^{(1)}_{\rm{21}}(\textbf{x},z) + 
\frac{1}{2}~\delta T^{(2)}_{\rm{21}}(\textbf{x},z),
\end{equation}
where $\TBRIGHT^{(0)}(z)\equiv\deltat$. Hence, $\Delta T_{21}(\textbf{x},z)$ can be expressed as 

\begin{eqnarray}
\Delta\TBRIGHT(\mathbf{x},z) &=& \TBRIGHT(\mathbf{x},z)-\deltat(z) \label{eq:T21_expansion}\\
\nonumber
&=& \mathcal{A}_{21}(z)\,\left( \delta^{(1)}(\mathbf{x},z)+ \frac{1}{2}\, \delta^{(2)}(\mathbf{x},z)\right) \nonumber\\
&+& \mathcal{B}_{21}(z)~\, \delta^{(1)}(\mathbf{x},z)^2 \label{eq:T21_expansion_tree}\\
&+& \mathcal{C}_{21}(z)\,\int d^3\mathbf{x_1} \int d^3\mathbf{x_2} \,\, \mathcal G_2(\mathbf{x_1},\mathbf{x_2},z) \nonumber\\
\qquad&& \delta(\mathbf{x}+\mathbf{x_1},z) \, \delta(\mathbf{x}+\mathbf{x_2},z) \;,
\nonumber
\end{eqnarray}
with $\delta^{(n)}$ indicating the $n$-th order fluctuation in baryon density. 
Note that throughout this work our analysis is restricted to the tree-level 21-cm power spectrum, in contrast to \cite{Pillepich:2006fj}. Consequently, we retain only terms up to first order in the density perturbation, \textit{i.e.}\ $\delta^{(1)}$, and neglect all higher-order contributions in Eq.~(\ref{eq:T21_expansion_tree}).
The coefficients in the expressions ($\mathcal{A}_{21}, \mathcal{B}_{21}, \mathcal{C}_{21}$) denote the couplings of the brightness temperature with baryon density (indicated by superscript $b$) and gas temperature (indicated by superscript $T$) and that's why these coefficients can be expressed as the weighted sum of the two contributions as:
\begin{eqnarray}
    \label{eq:f1}
    \mathcal{A}_{21}&=&A^{\rm{b}}_{21}+g_1~A^{\rm{T}}_{21},\\
    \label{eq:f2}
    \mathcal{B}_{21}&=&B^{\rm{bb}}_{21}+g^2_1~B^{\rm{TT}}_{21}+g_1~B^{\rm{bT}}_{21},\\
    \label{eq:f3}
    \mathcal{C}_{21}&=&A^{\rm{T}}_{21}.
\end{eqnarray}
$A^{\rm{b}}_{21}$ and $A^{\rm{T}}_{21}$ in Eq.~\eqref{eq:f1} can be expressed as following:
\begin{eqnarray}
\label{eq:A_21_b}
A^{\rm{b}}_{21}  &=& A^{\rm{0}}_{21}  \left[ \left( 1- \frac{\tgamma}{\mathcal{S}} \right) (1- \bar{x})\right.\nonumber\\ 
&-&\left.\frac{\tgamma~\mathcal{C}~(1+z)^3 } {\mathcal{Y}^2 ~\mathcal{S}^2} \Delta \mathcal{T} (1- \bar{x})^2  \right], \\
\label{eq:A_21_T}
A^{\rm{T}}_{21}  &=& A^{\rm{0}}_{21}  \left[ \frac{\tgamma~\mathcal{C}~(1+z)^3 } {\mathcal{Y} ~\mathcal{S}} \left( 1- 2~\frac{\Delta \mathcal{T}}{\mathcal{Y}~\mathcal{S}}\eta_1 \right)  (1- \bar{x})^2\right.\nonumber\\
&+& \left. \frac{\tgamma}{\mathcal{Y} ~\mathcal{S}^2} (\YA~\tk)(1- \bar{x}) \right], 
\end{eqnarray}
with
\begin{eqnarray}
\label{eq:A0_21_b}
    A^{\rm{0}}_{21} \simeq 69.05\left(\frac{\Omega_b^{(0)}h}{0.035}\right)\left(\frac{\Omega_m^{(0)}}{0.27}\right)^{-\frac{1}{2}}\left(\frac{1+z}{51}\right)^{\frac{1}{2}}\,{\rm mK}.
\end{eqnarray}
Here we have used the following quantities in order to determine Eqs.~\eqref{eq:A_21_b}-\eqref{eq:A0_21_b}:
\be
 \mathcal{C}= \phantom{a} \frac{4 ~\kappa_{\rm{10}}(\tk) ~T_{*}}{3~ A_{\rm{10}} ~ \tk}~  \bar{n}^0_{\rm HI}\;,
\ee
\be
 \mathcal{Y}_{\rm{C}}= \mathcal{C}(1+z)^3 (1-\bar{x})\;,\;\;
 \mathcal{Y}= \phantom{a}1 + \mathcal{Y}_{\rm{C}}\;,
\ee
\be
 \Delta \mathcal{T}=\phantom{a} \tgamma  - \tk\;,\;\;
 \mathcal{S}=\phantom{a}\frac{1}{\mathcal{Y}} \left[ \tgamma +  \mathcal{Y}_{\rm{C}} \tk  \right]\;.
\ee
Hence, using Eqs.~\eqref{eq:A_21_b}-\eqref{eq:A0_21_b}, we can calculate $\mathcal{A}_{21}(z)$ using Eq.~\eqref{eq:f1} where $g_1$ is calculated using a fitting form, given in \cite{Pillepich:2006fj}. However, unlike \cite{Pillepich:2006fj}, we have not considered the contributions from Lyman-$\alpha$ as we are interested to estimate $\mathcal{A}_{21}(z)$ at the Dark Ages.

\end{document}